\documentclass[12pt]{article}
\usepackage{amssymb,epsf,color}
\def\lsim{\mathrel{\rlap {\raise.5ex\hbox{$ < $}}
{\lower.5ex\hbox{$\sim$}}}}
\def\gsim{\mathrel{\rlap {\raise.5ex\hbox{$ > $}}
{\lower.5ex\hbox{$\sim$}}}} 
\def\sqr#1#2{{\vcenter{\vbox{\hrule height.#2pt

        \hbox{\vrule width.#2pt height#1pt \kern#1pt

           \vrule width.#2pt}

        \hrule height.#2pt}}}}

\def\lsim{{\displaystyle
{{\raise-8pt\hbox{$ <$}}
\atop{\raise5pt\hbox{$\sim$}}}}}
\def\gsim{{\displaystyle
{{\raise-8pt\hbox{$ >$}}
\atop{\raise5pt\hbox{$\sim$}}}}}
\def\slsim{{\displaystyle
{{\raise-8pt\hbox{$\scriptstyle <$}}
\atop{\raise5pt\hbox{$\scriptstyle \sim$}}}}}
\def\sgsim{{\displaystyle
{{\raise-8pt\hbox{$\scriptstyle  >$}}

\atop{\raise5pt\hbox{$\scriptstyle \sim$}}}}}

\newskip\humongous \humongous=0pt plus 1000pt minus 1000pt

\newcommand{\sumpf}[0]{\sum_{(H^{\rm f},G^{\rm f})}\! \! \! \!
{\raise
4pt
\hbox{$'$}}\,}

\newcommand{\sump}[0]{\sum_{(H,G)}\! \! {\raise 4pt \hbox{$'$}}\,}

\def\bs{\begin{subequations}}
\def\es{\end{subequations}}

\catcode`\@=11
\newcount\hour
\newcount\minute
\newtoks\amorpm

\hour=\time\divide\hour by60
\minute=\time{\multiply\hour by60 \global\advance\minute by-\hour}
\edef\standardtime{{\ifnum\hour<12 \global\amorpm={am}%
        \else\global\amorpm={pm}\advance\hour by-12 \fi

        \ifnum\hour=0 \hour=12 \fi
        \number\hour:\ifnum\minute<10 0\fi\number\minute\the\amorpm}}
\edef\militarytime{\number\hour:\ifnum\minute<10 0\fi\number\minute}
\def\draftlabel#1{{\@bsphack\if@filesw {\let\thepage\relax
   \xdef\@gtempa{\write\@auxout{\string
      \newlabel{#1}{{\@currentlabel}{\thepage}}}}}\@gtempa
   \if@nobreak \ifvmode\nobreak\fi\fi\fi\@esphack}
        \gdef\@eqnlabel{#1}}
\def\@eqnlabel{}
\def\@vacuum{}
\def\draftmarginnote#1{\marginpar{\raggedright\scriptsize\tt#1}}
\def\draft{\oddsidemargin -.2truein
        \def\@oddfoot{\sl preliminary draft \hfil
        \rm\thepage\hfil\sl\today\quad\militarytime}
        \let\@evenfoot\@oddfoot \overfullrule 3pt
        \let\label=\draftlabel
        \let\marginnote=\draftmarginnote
   \def\@eqnnum{(\theequation)\rlap{\kern\marginparsep\tt\@eqnlabel}%
\global\let\@eqnlabel\@vacuum}  }
\def\subequations{\refstepcounter{equation}%
  \edef\@savedequation{\the\c@equation}%
  \@stequation=\expandafter{\theequation}
  \edef\@savedtheequation{\the\@stequation}
  \edef\oldtheequation{\theequation}%
  \setcounter{equation}{0}%
  \def\theequation{\oldtheequation\alph{equation}}}

\def\endsubequations{\setcounter{equation}{\@savedequation}%
  \@stequation=\expandafter{\@savedtheequation}%
  \edef\theequation{\the\@stequation}\global\@ignoretrue
  \vspace*{-12pt} \\}

\def\bs{\begin{subequations}}
\def\es{\end{subequations}}
\relax
\def\thefootnote{\fnsymbol{footnote}}
\def\be{\begin{equation}}
\def\ee{\end{equation}}
\def\ba{\begin{eqnarray}}
\def\ea{\end{eqnarray}}

\def\ee{\end{equation}}
\def\bea{\begin{eqnarray}}
\def\eea{\end{eqnarray}}
\def\nn{\nonumber}

\newcommand{\uarrw}[0]{\mathrel{
{\raise.5ex\vbox{\hrule width 1cm}\hskip-6pt\rightarrow}}}
\def\thebibliography#1{%
\vskip 0.5cm \centerline{\bf References}
\list{%
[\arabic{enumi}]}{\settowidth\labelwidth{[#1]}
\leftmargin\labelwidth
\advance\leftmargin\labelsep
\usecounter{enumi}}
\def\newblock{\hskip .11em plus .33em minus .07em}
\sloppy\clubpenalty4000\widowpenalty4000
\sfcode`\.=1000\relax}

\renewcommand{\theequation}{\arabic{section}.\arabic{equation}}

\renewcommand{\section}{\setcounter{equation}{0}\@startsection%
{section}{1}{0mm}{-\baselineskip}{0.5\baselineskip}%
{\normalfont\normalsize\bfseries}}

\renewcommand{\subsection}{\@startsection%
{subsection}{2}{0mm}{-\baselineskip}{0.5\baselineskip}%
{\normalfont\normalsize\slshape}}

\renewcommand{\subsubsection}{\@startsection%
{subsubsection}{2}{0mm}{-\baselineskip}{0.5\baselineskip}%
{\normalfont\normalsize\slshape}}

\begin{document}
%
%
\renewcommand{\theequation}{\arabic{section}.\arabic{equation}}
\begin{titlepage}
\begin{flushright}
\end{flushright}
\begin{centering}
\vspace{1.0in}
\boldmath

{ \large \bf Is the Universe the only existing Black Hole? \\
\bf }

\unboldmath
\vspace{1.5 cm}

{\bf Andrea Gregori}$^{\dagger}$ \\
\medskip
\vspace{3.2cm}
{\bf Abstract} \\
\end{centering} 
\vspace{.2in}
We investigate the physics of 
black holes in the light of the quantum theoretical framework proposed 
in~\cite{assiom}. It is argued that 
black holes are completely non-local objects, and that 
the only one which really exists 
is the universe itself.

\vspace{9cm}

\hrule width 6.7cm
\noindent
$^{\dagger}$e-mail: agregori@libero.it

\end{titlepage}
\newpage
\setcounter{footnote}{0}
\renewcommand{\thefootnote}{\arabic{footnote}}

\section{Introduction}

Black holes are ``singularities'' of the space-time, that,
according to the Einstein's theory of Relativity, occur
when the matter/energy density contained within a region of space exceeds 
a bound, given by the Schwarzschild relation:
\be
{2GM \over c^2} \, = \, R
\label{schw}
\ee 
Black holes are expected to exist and probably lie at the center
of every Galaxy. Only in this way it seems possible to  
justify the high gravitational attraction implied by
the observed orbital speed of stars in the part of the coil near the center.
However, is it really possible to localize such a 
huge concentration of mass/energy
within a well defined sub-region of the universe? From a classical 
point of view there is no problem to give a positive answer. 
The situation may however be rather different from a quantum point of view.
Indeed, although black holes are introduced as ``classical'' objects, 
predicted by the theory of Relativity,
we know that, at a certain scale, the physical world
shows out its quantum mechanical nature. This is expected to occur also for 
black holes.
The quantum physics of black holes has been
the matter of several investigations, 
in more recent times also within the context of string theory.
Fundamental in this respect are the studies carried out by J. Bekenstein and
S. Hawking,
leading to the evaluation of the black holes entropy as a function
of the area of the horizon~\cite{bekenstein,hawk1}, 
and to the prediction of their
evaporation~\cite{hawk2}. 

In Refs.~\cite{assiom,rel} I proposed a theoretical framework based on few 
fundamental assumptions, that produce a physical scenario which embeds both
Quantum Mechanics and Relativity, to which it reduces in appropriate limits. 
However, it is constructed by assuming no one of the usual properties at the 
ground of either one of these theories. The way it reduces to the one or the 
other resembles a bit the way quantum mechanics "reduces" to classical 
mechanics in the limit $\hbar \to 0$.
That is, there is indeed no real smooth limit in the mathematical sense, the 
sense in which a term of a Lagrangian, or something alike, which parametrizes 
the deviation from the classical theory, goes to zero. Quantum mechanics can 
only be formally put in the form of a $\hbar$-dependent correction to 
classical mechanics. Indeed, it entails a completely different approach to 
causality, temporal evolution, etc... In a similar way, the 
approach introduced in \cite{assiom}
is well approximated by Quantum Mechanics, or Relativity, under certain 
conditions, without being (an extension of) either of them in a strict sense. 

As the resulting scenario has proven to be compatible with all current 
experimental results for what concerns the physics of elementary particles and 
several basic aspects of cosmology and the evolution of the 
universe~\cite{spi}, 
it is interesting to investigate within this theoretical framework also the 
physics of black holes. 
Namely, to see what are the implications for the physics of black holes of a 
relativistic-quantum mechanical scenario, i.e. a quantum gravity scenario,
in which quantum mechanics is not dealt with, as usual, as an ``ad hoc'' 
modification of the rules of classical mechanics, but is a necessary and 
fundamental theoretical implication.

\newpage

\section{The combinatorial scenario}

In Ref.~\cite{assiom} I proposed that our quantum world can be viewed as
the result of the superposition of any possible configuration. By configuration
it is here meant any assignment/distribution of ``energy'' along a target
space, that can be of whatever number of dimensions. 
Energy too is intended in the most 
elementary sense. One could also speak of ``units of curvature'', but
more fundamentally it is only an assignment of a binary code of ``occupation''
to a target space of ``unoccupied'' positions. 
At any fixed amount $E$ of ``energy units'', the ``partition function'' of the 
universe, i.e. the generating function for
the mean value of any observable, of the universe, is given by
the sum over all such configurations $\psi_E$, weighted with their
volume of occupation in the phase space of all the configurations, 
$W(\psi_E) = \exp S(\psi_E)$:
\be
{\cal Z}_E ~ = ~ \int {\cal D} \psi_E {\rm e}^{S(\psi_E)} \, .
\label{Zsum}
\ee 
Only in an average sense, and 
only once a limit to a continuum description is taken,
after the introduction of a unit of length and of energy, one can speak
of energy, space, curvature in the ordinary sense. 
As nothing basically distinguishes the nature of the elements of the ``energy''
space from those of the ``positions space'', apart from being the 
ones working as target and the others as ``base'' space, in some sense 
the set $\left\{  \psi_E \right\}$ of all the maps at given $E$ can be viewed  
as the space of all possible structures one can think about, all possible 
assignments from a space to another one. 
The energy $E$ turns out
to play the role of a label for an ordering, given by the inclusion of
phase spaces, $\left\{ \psi_E  \right\} \subseteq \left\{ \psi_{E^{\prime}}  
\right\}$ if $E \leq E^{\prime}$. In the continuum limit, $E$ plays 
therefore also the role of time, or age of the universe.
The fact that 
concepts like energy, space, time, curvature, 
are only ``large scale'' and \underline{mean} quantities, leads to an 
indeterminacy of observables, whose mean values are smeared, unfocused, by an
amount that turns out to correspond to the Heisenberg's 
uncertainty of quantum mechanics.
Indeed, the latter can be viewed as the implementation of this 
uncertainty principle, obtained through 
a scenario of waves and probabilities. 

As discussed in Ref.~\cite{assiom}, \ref{Zsum} implies that the appearance of
the universe is dominated by the most entropic configurations; in these
configurations the space is three dimensional, with the curvature of a 
three-sphere, whose radius is given by the total energy/age of the universe
$E$. 
One can then show that the speed of expansion of the
--~average, three-dimensional~-- universe, that by convention and choice of
units we can call ``$c$'', is also the
maximal speed of propagation of \underline{coherent}, i.e.
non-dispersive, information. In the
limit in which one passes to the continuum and speaks of space, namely
when one speaks of average three-dimensional world, this can be shown
to correspond to the $v=c$ bound of the speed of light\footnote{Here it is
essential that we are talking of coherent information, as tachyonic
configurations also exist in this scenario, which embeds also Quantum
Mechanics.}. Moreover, the geometry of geodesics in this space corresponds to
the one generated by the energy distribution. This means that this
framework ``embeds'' in itself Special and General Relativity \cite{rel}.

The theoretical framework proposed in \cite{assiom} goes therefore beyond 
both quantum mechanics and the theory of relativity, lifting them to a 
description which is fundamentally neither quantum mechanical nor relativistic,
and therefore not quantum-field-theoretical either. 
These theories constitute good approximations of it in appropriate limits.

Relevant for our present discussion is that, in particular, this scenario 
allows us to deal with coordinate transformations, and therefore also with
the metric, in a generalised sense, beyond the usual distinction between
relativistic (classical geometrical) 
and quantum mechanical aspects. There is no
more ``classical theory'' which is going to be quantised, by applying a 
``quantum suit'' to a basically classical description. To better appreciate
the fundamental difference of the two approaches, one must consider that,
when quantizing a classical system, certainly one modifies the rules of the
classical game, but, in some sense, he works on an already ``projected out'' 
system, which
has been first reduced to classical terms, and then 
``theoretically expanded'' through
a quantization procedure. Already thinking in terms of space, and a theory  
of quantum fields on it, is such a kind of ``two steps procedure'',
which can be misleading in some cases. This is particularly
true when quantum effects
eventually destroy the classical sense of space and time. 
Black holes are an example of system in which indeed these concepts
are pushed to their natural limits of definition already within
the classical theory. Applying a ``quantization procedure'' to such
a critical situation may be not appropriate.
The framework introduced in \cite{assiom} provides us with
a direct way of dealing with observables without passing
through a classical description of physics.

\section{The metric around a black hole}

Space, and metric, are average concepts that arise 
only at a relatively ``large'' scale. As a consequence, this is true also
for coordinate transformations, and the metric of space.  
At a more microscopical level, i.e. at a shorter length scale
(and therefore also in a deeply quantum regime),  
they must be substituted by expressions relating  
the variation of entropy, as it is perceived by different observers \cite{rel}.
The metric of space-time precisely arises as large scale limit of
a quantity that expresses the rate of local variation of entropy.
Entropy is the quantity that substitutes, at a more fundamental
level, time variation and curvature of space.
The general expression relating the evolution of a system as is seen
by the system itself, that we indicate with $A^{\prime}$,
and by an external observer, $A$, is given, according to
\cite{rel}, by:
\ba
\Delta S(A) & = & \Delta S ({\rm internal}\, = \, {\rm at \, rest}) 
\, + \, \Delta S ({\rm external})  \label{DeltaSS} \\
&& \nn \\
& = & \Delta S (A^{\prime}) \, + \, 
\Delta S_{\rm external}(A) \, ,  
\label{DeltaSSk}
\ea
where $\Delta S(A)$ on the left hand side
is the variation of entropy of the event which is 
detected, as seen from the observer $A$, 
whereas on the right hand side $\Delta S (A^{\prime})$
is the variation of entropy as seen from the system itself, $A^{\prime}$.
$\Delta S ({\rm external})$ is the difference between the two, namely, the
amount of entropy variation that $A$ refers to the environment of $A^{\prime}$,
and not to something ``built in'' in $A^{\prime}$. For instance, 
$\Delta S ({\rm external})$ is the effect of an external force, or 
the variation of entropy due to the motion itself of the frame comoving
with $A^{\prime}$ (see Ref.~\cite{rel} for more detail and explanations).

The classical limit of a physical system corresponds to the limit in which
the scale is sufficiently large to enable not only talking of smooth geometry,
smooth coordinates like space, energy, time, 
instead of simple combinatorials of distributions of energy,
but also to make possible
considering the average, mean values of observables to be well approximated
by the dominant, most entropic configurations of the universe. More remote
configurations build up the ``quantum fluctuation'' around classical values.
Deeply quantum mechanical systems are those in which the contribution of
more remote configurations is no more negligible.

In the large-scale, \emph{classical} limit, the variations of entropy
$\Delta S(A)$ and $\Delta S (A^{\prime})$ can be written in terms
of time intervals: 
\be
\Delta S(A) \rightarrow \langle \Delta S(A)  \rangle 
\approx (c \Delta t)^2 \, ,
\label{dst}
\ee
and
\be
\Delta S (A^{\prime}) \rightarrow 
\langle \Delta S (A^{\prime})  \rangle
\approx (c \Delta t^{\prime})^2 \, ,
\label{dstp}
\ee
where we omitted
universal proportionality constants (from now on, we will
also omit the speed of light $c$, that we set to 1, as we also implicitly
did for the Boltzmann constant, and all other fundamental scales and 
constants).
$t$ and $t^{\prime}$ are respectively
the time as measured by the observer, and the proper time of the system
$A^{\prime}$.
In this case, expression~\ref{DeltaSSk} can be written as:
\be
(\Delta t^{\prime})^2 ~= ~ (\Delta t)^2 \, - \, 
\langle \Delta S^{\prime}_{\rm external}(t) \rangle \, ,
\label{SpSS}
\ee
The temporal part of the metric is therefore given by:
\be
g_{00} ~ = ~ 
{\langle \Delta S^{\prime}_{\rm external}(t) \rangle 
\over (\Delta t)^2} \, - \, 1   \, . 
\label{g00}
\ee
As long as we consider systems for which $g_{00}$ is far
from its extremal value, expression \ref{g00} constitutes a good approximation
of the time component of the metric. 
However, a black hole does not fall within the domain of this
approximation.
According to its very (classical) definition, 
the only part we can probe of a black hole is the surface at the horizon.
In the classical limit the metric at this surface vanishes:
$g_{00} \to 0$ (an object falling from outside toward the black hole
appears to take an infinite time in order to reach the surface).
This means,
\be
\langle \Delta S_{\rm external} \rangle
~ \approx ~\propto \, \left( \Delta t  \right)^2 \, . 
\ee 
However, in our set up time is only an average, ``large
scale'' concept, and only in the large scale, classical limit we
can write variations of entropy in terms of progress of a time coordinate
as in \ref{dst} and \ref{dstp}.
The fundamental transformation is the one given in expressions 
\ref{DeltaSS}, \ref{DeltaSSk}, and the term $g_{00}$ has only to be understood
in the sense of:
\be
\Delta S (A^{\prime}) \, \longrightarrow \,  
\langle \Delta S (A^{\prime}) \rangle \, \equiv \, 
\Delta t^{\prime} g_{00} \Delta t^{\prime} \, . 
\label{dsag}
\ee
As discussed in Ref.~\cite{rel},
the apparent vanishing of the metric~\ref{g00}
is due to the fact that we are subtracting contributions from the
first term of the r.h.s. of expression~\ref{DeltaSSk}, namely 
$\Delta S(A^{\prime})$,
and attributing them to the contribution of the
environment, the world external to the system of which we consider the proper 
time, the second term in the r.h.s. of \ref{DeltaSSk}, 
$\Delta S_{\rm external}(A)$. 
Any physical system is given by the superposition of an
infinite number of configurations, of which only the most entropic ones
(those with the highest weight in the phase space) build up
the classical physics, while the more remote ones contribute to 
what we globally call ``quantum effects''. Therefore,
taking out classical terms from
the first term, $\Delta S (A^{\prime})$, the ``proper frame'' term,
means transforming the system the more and more into a ``quantum
system''. In particular, this means that the mean value of whatever observable
of the system will receive the more and more contribution 
by less localized, more exotic, configurations, 
thereby showing an increasing quantum uncertainty.
In particular, the system moves toward configurations 
for which $\Delta x \rightarrow \gg 1 / \Delta p$.
Indeed, one never reaches the condition of vanishing of \ref{dsag}, because,
well before this limit is attained, also the notion itself of space, and time,
and three dimensions, localized object, geometry, etc..., are lost. The
most remote configurations in general do not
describe a universe in a three-dimensional space, and the ``energy'' 
distributions are not even interpretable in terms of ordinary observables
(see discussion in~\cite{assiom,rel}).
At the limit in which we reach the surface of the horizon, the black hole will
therefore look like a completely delocalized object.

\begin{figure}
\centerline{
\epsfxsize=10cm
\epsfbox{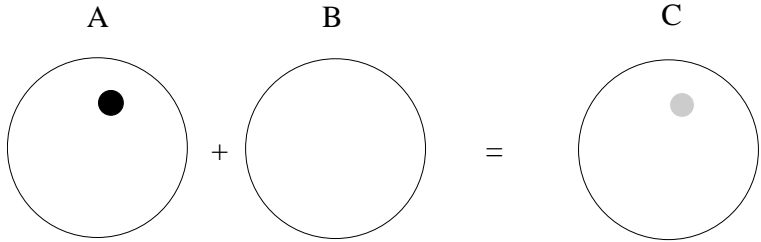}
}
\vspace{0.3cm}
\caption{The superposition of two configurations of the universe: $A$, 
which has a region of mass density corresponding to a black hole,
here coloured in black, and $B$, where this region is absent. If we represent 
the amount of mass density through tones of grey, so that black is the 
critical black hole mass, and white is zero mass (or, more precisely, the
ground energy density of the universe, $\Lambda$), we can represent the 
superposition of $A$ and $B$ as $C$, where the black hole region of $A$ is
``softened'' to a grey coloured region, no more a black hole. 
The tone of grey is
the more and more lighter, the smaller is the weight of
configurations like $A$ as compared to configurations of the type $B$ in
the phase space of all the configurations. Saying that around a black hole
physics is in a highly quantum, delocalized regime, means precisely 
that configurations like $A$ weight much less than those like $B$, or $C$.}
\label{blackhole-bw}
\end{figure}

According to \ref{Zsum} the universe that one observes is the 
superposition of all its possible configurations. 
In this theoretical framework, the existence of a black hole
as a localized object within the universe does not
simply mean that there exist configurations
in which a concentration of mass with the characteristics of a black hole:
this is obviously true, because \ref{Zsum} sums over \emph{all} the
configurations one can think about.
In order to have a localized black hole it is also necessary 
that these configurations contribute to \ref{Zsum} with a 
sufficiently large weight (i.e. they must be sufficiently entropic in
the phase space of all the configurations of the universe). Otherwise,
if such configurations are only ``remote'', the averaged effect of the 
superposition with configurations which don't show such a concentration of mass
results in the spoiling of the black hole, the effective removal of
the Schwarzschild singularity. Saying that the horizon of a
well localized black hole
belongs to a class of configurations which are remote in the phase space
precisely means that, in the resulting universe, such a black hole in practice
does not exist (see figure~\ref{blackhole-bw}).    

\section{The universe itself as a black hole}

\begin{figure}
\centerline{
\epsfxsize=7cm
\epsfbox{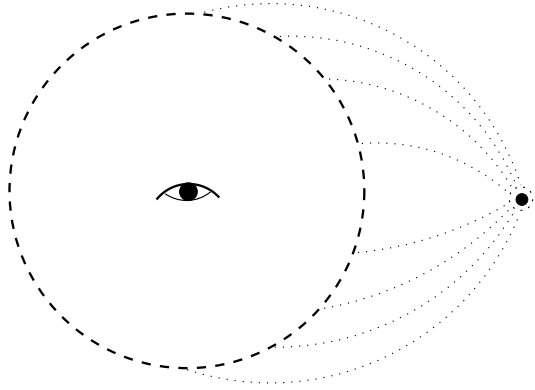}
}
\vspace{0.3cm}
\caption{The universe, at the center of which we, the observer, sit.
The dashed horizon, apparently a classical surface of area ${\cal T}^2$,
from a physical point of view corresponds to a point, the point at the origin. 
A ``quantum delocalized'' point.}     
\label{pointhoriz}
\end{figure}

Saying that an object is completely delocalized is like saying that it is
extended as the universe itself. 
In Refs.~\cite{spi,assiom} we indeed spoke of the universe as a black hole.
In the theoretical framework we are considering, 
the universe is \emph{classically} extended up to the horizon 
corresponding in light years to its age, $R = c {\cal T}$, and has a total
energy also proportional to its age/radius, $E \propto R ( \propto {\cal T})$.
The energy density of the universe is of order
$\rho \sim 1/{\cal T}^2$ for any of the three types of energy density
(cosmological, matter and radiation densities).
In particular, this is true also for the cosmological constant: 
$\Lambda \sim 1/{\cal T}^2$. 
At the present, the latter
corresponds to $H^2$, the Hubble parameter to the square.
However, the relation $\rho \sim 1/{\cal T}^2$, a pure experimental numerical 
observation of the present time universe, in our set up is promoted
to a general functional dependence of the energy densities on the age of the 
universe.
Since the universe is trivially ``at rest'', 
its total energy coincides with its 
``rest mass''; the relation between energy and radius of the
universe corresponds therefore to the Schwarzschild relation 
for the radius of a black hole. 
Indeed, a black hole doesn't need to have an extremely high mass density, 
because the relation
\ref{schw} only states a proportionality of mass and radius of a spheric 
region of space. As the radius increases, the mass density decreases
like $\sim 1/R^2$, and the black hole
becomes the more and more rarefied.
There is nothing odd in a universe behaving like a black hole.
The universe is non-observable from outside, 
because of the simple fact that there is no outside of it. 
As there is no outside of the universe, there is also trivially
no information ``going out'' from it. Or, if one prefers, information expands
comovingly with the horizon of the universe, i.e. at speed $c$, the speed of 
light. Therefore, the forefront of the wave carrying information is always 
``stuck'' on the horizon. 
Light rays don't travel across the horizon, 
therefore don't go out of the universe, for the simple fact that they ``stir''
the horizon itself. In our scenario, light rays
are the more and more red-shifted,
the closer and closer they are to the classical horizon of the 
universe, till
a limit of infinite wavelength \footnote{The closer and closer to the horizon,
radiation emitted by matter
gets on the other hand a violet-shift that partially counters the red-shift 
effect. For a discussion of this, and also the consequent apparent acceleration
of the expansion of the Universe, see \cite{spi}.}, as it happens
in the case of a black hole, as seen both from the outside and from the 
inside.   
On the other hand, infinite wavelength does not mean that light
employs an infinite time for travelling from the horizon to us:  
the universe is a black hole in expansion at the speed of light, and 
the horizon is at a distance corresponding in light years to the age of
the universe. We remark that a quantum 
delocalization of any point at the horizon of the universe is precisely what
we need in order to resolve the apparent paradox originating
from the fact that, in this interpretation, the surface at the horizon 
corresponds to a (Planck size) point, the origin of the universe
\footnote{See discussion of section~2 of
Ref.~\cite{spi}.} (see figure~\ref{pointhoriz}). 

\vspace{.5cm}

According to our considerations, the only
black hole in the whole universe is the universe itself, trivially non-local,
non-observable from outside, and with all the characteristics of mass, radius,
shifts of frequencies typical of a black hole.
Our conclusions are therefore quite far away from the ones derived within
a traditional quantum analysis of the physics of a classical black hole.
The only result we have in common with the traditional quantum mechanical
analysis is about the black holes entropy, given in terms of the
area of the surface at the horizon. In our case, the only 
physical system to which this concretely applies is however 
the universe itself. 
The scaling of
the entropy like the square of the radius/time/energy is derived
from statistical considerations, as
a consequence of the fact that
in our scenario the universe turns out to possess in the average the
geometry of a three-sphere (see Ref.~\cite{assiom}, section~7).

\section{Black holes at the center of galaxies?}

Observations made on the orbital speed of stars relatively close 
to the center of some galaxies indicate the typical behaviour of a body
subjected to a very strong gravitational force.
This has induced to suspect the existence of a black hole possibly 
at the center of every galaxy \cite{schoedel,ghez,gerssen,peplov}.
Indeed, the angular velocity $\omega$ of an orbiting object is related to the
radius $R$ of the orbit and the mass $M$ of the center \footnote{We assume here
for simplicity the mass of the center of the galaxy to be much larger 
than the mass of the 
orbiting star, so that we can approximate the reduced mass with the mass of 
the star, and the mass of the center of mass with the mass of the center of 
the galaxy.} by the well known expression:
\be
{1 \over 2} \omega^2 R^2 ~ = ~ {M \over R} \, .
\label{orbit}
\ee
This relation is in general valid only pointwise along an elliptical orbit,
but for the present discussion we can even assume to work with circular ones. 
Here it is important to point out what is experimentally measured, and what is 
derived through an interpretation of experimental data within a theoretical
framework. The key experimental observation is
the \underline{period} of the orbit, from which one derives the angular
velocity $\omega$. The mass $M$ of the supposed-to-be black hole is then
derived after the measurement of the radius $R$ of the orbit.

According to the cosmological scenario resulting from our theoretical 
framework, distances in regions of space corresponding to the past of our
universe appear the more and more expanded, as we approach the horizon
of observation.
This means that the orbital lengths in galaxies which are far away from
us appear larger than what they indeed are (see for instance
the discussion in section~9 of Ref.~\cite{spi}).
Measured lengths must be contracted in order to obtain the real ones. 
By looking at expression~\ref{orbit} one can see that, at fixed
$\omega$, the mass $M$ scales with $R^3$. This means that, if
the real orbital radius is a factor $K < 1$ smaller than the
observed one, the mass $M$ of the center of the galaxy is indeed
a factor $K^3 \ll 1$ smaller than what inferred within the usual theoretical
schemes. Since the Schwarzschild radius scales linearly with the black hole
mass, the rescaling between apparent and real lengths leads us a factor
$1/K^{2} \gg 1$ far away from the critical mass/radius threshold
that would induce one to expect the presence of a black hole at the center of 
the galaxy. In other words, not only the observed orbital periods can be 
justified with much smaller central masses, but the scaling
relation is such that masses decrease much faster than radii, in such a way
that they remain much further below the critical black hole mass density.

\vspace{1.5cm}

\providecommand{\href}[2]{#2}\begingroup\raggedright\endgroup

\end{document}